\documentclass[twoside,11pt]{article}
\usepackage[preprint]{jmlr2e}
\usepackage[utf8]{inputenc}
\usepackage[T1]{fontenc}
\usepackage{listings}
\usepackage{amsfonts}
\usepackage{amsopn}
\usepackage{amsmath}
\usepackage{graphicx}
\usepackage{color}
\usepackage{url}
\usepackage{enumitem}
\usepackage{lastpage}

\usepackage{hyperref}
\hypersetup{
  colorlinks=true,
  linkcolor=dark-blue,
  urlcolor=dark-blue,
  citecolor=dark-blue,
}

\definecolor{dark-blue}{rgb}{0.15,0.15,0.4}
\definecolor{codegreen}{rgb}{0,0.6,0}
\definecolor{codegray}{rgb}{0.5,0.5,0.5}
\definecolor{codepurple}{rgb}{0.58,0,0.82}
\definecolor{backcolour}{rgb}{0.95,0.95,0.92}

\setitemize{noitemsep}
\setenumerate{noitemsep}

\definecolor{mygreen}{rgb}{0,0.6,0}
\definecolor{mygray}{rgb}{0.5,0.5,0.5}
\definecolor{mymauve}{rgb}{0.58,0,0.82}

\jmlrheading{12}{2025}{1-\pageref{LastPage}}{x/xx; Revised x/xx}{x/xx}{21-0000}{Oleg Smirnov}

\ShortHeadings{\texttt{tensorflow-riemopt}: A Library for Optimization on Riemannian Manifolds}{Smirnov}
\firstpageno{1}

\title{\texttt{tensorflow-riemopt}: A Library for Optimization on Riemannian Manifolds}

\begin{document}

\author{\name Oleg Smirnov \email oleg.smirnov@microsoft.com \\ \addr Microsoft, Stockholm, Sweden
}

\editor{}

\maketitle

\begin{abstract}%
This paper presents \texttt{tensorflow-riemopt}, a Python library for 
geometric machine learning in TensorFlow~\citep{abadi2016tensorflow}. 
The library provides efficient implementations of neural network layers with 
manifold-constrained parameters, geometric operations on Riemannian 
manifolds, and stochastic optimization algorithms for non-Euclidean 
spaces. Designed for integration with TensorFlow Extended, it supports 
both research prototyping and production deployment of machine learning 
pipelines. The code and documentation are distributed under the MIT 
license and available at \url{https://github.com/master/tensorflow-riemopt}
\end{abstract}


\section{Introduction}

Over the past years there has been a surge of interest in machine learning in non-Euclidean domains. Representation learning in spaces of constant negative curvature, that is, hyperbolic, has shown to outperform Euclidean embeddings significantly on data with latent hierarchical, taxonomic or entailment structures. Such data naturally arises in language modeling~\citep{nickel2017poincare, nickel2018learning, chamberlain2017neural, sala2018representation, ganea2018hyperbolic}, recommendation systems~\citep{chamberlain2019scalable}, image classification, and few-shot learning tasks~\citep{khrulkov2020hyperbolic}. Matrix manifolds, such as Grassmannian and Stiefel spaces, find applications in computer vision to perform emotion and action recognition from video data~\citep{chakraborty2019statistics, huang2017riemannian}, face recognition and image set classification~\citep{huang2018building, li2015face}. Manifolds of symmetric positive definite (SPD) matrices characterize data from diffusion tensor imaging~\citep{pennec2006riemannian} and functional magnetic resonance imaging~\citep{sporns2005human}. The Lie groups of transformations $\mathrm{SO}(3)$ and $\mathrm{SE}(3)$ appear when dealing with problems like pose estimation and skeleton-based action recognition~\citep{hou2018computing, huang2017deep}. Among those advances is a family of neural network methods, which are parameterized by weights constrained to a particular non-Euclidean space.

Remannian optimization algorithms including adaptations of stochastic gradient descent (SGD) methods to non-Euclidean settings~\citep{bonnabel2013stochastic, roy2018geometry, becigneul2018riemannian, kasai2019riemannian} have been robustly integrated with many popular machine learning toolboxes. However, previous work in this space was mainly motivated by research use cases~\citep{townsend2016pymanopt, meghwanshi2018mctorch, miolane2020geomstats, kochurov2020geoopt}, whereas practical aspects, such as deploying and maintaining machine learning models, were often overlooked. The need for production-ready infrastructure is evidenced by recent applications such as hyperbolic modeling in search ranking~\citep{choudhary2022anthem}, which required a large-scale deployment. We present \texttt{tensorflow-riemopt}, a Python library for optimization on Riemannian manifolds in TensorFlow~\citep{abadi2016tensorflow} to help bridge the aforementioned gap. The library targets not only research, but also production use cases with the help of integration with the TensorFlow Extended platform~\citep{baylor2017tfx}.

\section{Usage modes}
\texttt{tensorflow-riemopt} supports two usage modes for machine learning applications. The first enables practitioners to deploy geometric ML models in production, while the second provides researchers with building blocks for developing custom algorithms.
 
\paragraph{Machine learning applications.} For practitioners deploying geometric machine learning models, the library integrates with the TensorFlow ecosystem through manifold-constrained variables. Listing~\ref{lst:nn_usage} illustrates a bilinear mapping layer from SPDNet~\citep{huang2016riemannian}, where weight matrices lie on the Stiefel manifold.

Manifold-constrained variables can be transparently used with standard TensorFlow operations and training loops, enabling deployment through TensorFlow Extended pipelines. Native TensorFlow tensors are automatically treated as Euclidean data by Riemannian optimizers, allowing seamless mixing of manifold and standard layers. 

\begin{lstlisting}[language=Python,caption={Manifold-constrained layer},label={lst:nn_usage},basicstyle=\scriptsize]
from tensorflow_riemopt.variable import assign_to_manifold
from tensorflow_riemopt.manifolds import StiefelEuclidean
from tensorflow_riemopt.optimizers import RiemannianSGD

# Define layer with manifold constraint
class BiMap(tf.keras.layers.Layer):
    def build(self, input_shape):
        self.w = self.add_weight("w", shape=[input_shape[-1], self.output_dim])
        assign_to_manifold(self.w, StiefelEuclidean())  # Key operation
    
    def call(self, inputs):
        return tf.transpose(self.w) @ inputs @ self.w

# Train with Riemannian optimizer
model.compile(optimizer=RiemannianSGD(learning_rate=0.01), ...)
\end{lstlisting}

The repository provides example implementations of SPDNet for emotion recognition, GrNet~\citep{huang2018building} for video classification on Grassmann manifolds, and LieNet~\citep{huang2017deep} for skeleton-based action recognition.

\begin{figure}[t]
  \begin{minipage}{.5\textwidth}
    \centering
    \begin{lstlisting}[language=Python,caption={Riemannian geometry operations},label={lst:low_level_api},basicstyle=\scriptsize]
from tensorflow_riemopt.manifolds import Sphere

# Instantiate manifold
S = Sphere()
x = S.projx(tf.constant([0.1, -0.1, 0.1]))
u = S.proju(x, tf.constant([1., 1., 1.]))
v = S.proju(x, tf.constant([-0.7, -1.4, 1.4]))

# Apply geometric operations
y = S.exp(x, v)
u_ = S.transp(x, y, u)
v_ = S.transp(x, y, v)
    \end{lstlisting}
  \end{minipage}
  \begin{minipage}{.5\textwidth}
    \centering
    \includegraphics[width=\textwidth]{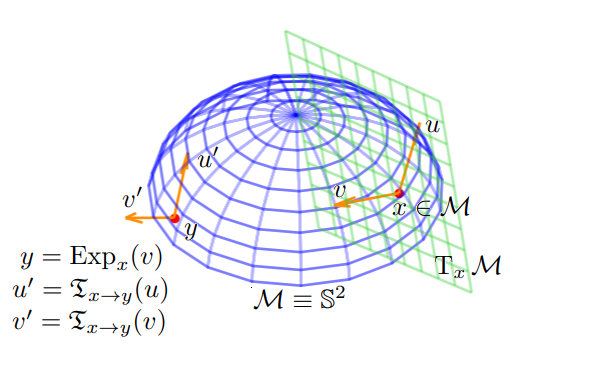}
    \caption{Geometric operations on $\mathbb{S}^2$}
    \label{fig:low_level_api}
  \end{minipage}
\end{figure}

\paragraph{Riemannian geometry operations.} For researchers developing custom algorithms or extending the library with novel manifolds, geometric operations are exposed directly through a low-level API. Listing~\ref{lst:low_level_api} demonstrates fundamental operations on the sphere manifold, including projections, exponential maps, and vector transports, with their interpretation shown in Figure~\ref{fig:low_level_api}.

This mode supports implementation of novel Riemannian optimization algorithms and custom manifolds. Both modes share the same efficient underlying implementations, enabling researchers to prototype algorithms with the low-level API and practitioners to deploy them through the high-level integration.

\section{Implementation details}

\texttt{tensorflow-riemopt} builds upon the TensorFlow framework~\citep{abadi2016tensorflow} and its automatic differentiation capabilities. The library is designed for seamless integration with the TensorFlow ecosystem, with all components serving as drop-in replacements for native APIs. This ensures compatibility with TensorFlow Extended~\citep{baylor2017tfx} in both graph and eager execution modes, supporting the full lifecycle of production machine learning pipelines. The implementation prioritizes closed-form expressions for manifold operators with numerical approximations as fallback, and supports efficient updates on dense and sparse tensors.

The package implements manifolds and Riemannian metrics with their associated exponential and logarithmic maps, geodesics, retractions, and vector transports. Following the Manopt~\citep{boumal2014manopt} design, each manifold is implemented as a Python class inheriting from an abstract \texttt{Manifold} base class. The minimal interface for optimization consists of four operations: the Riemannian metric \texttt{inner(x, u, v)}, tangent space projection \texttt{proju(x, u)}, retraction \texttt{retr(x, u)}, and vector transport \texttt{transp(x, y, v)}. The library currently supports 14 manifolds including hyperbolic spaces (Poincaré ball and hyperboloid models), matrix manifolds (Grassmannian, Stiefel with multiple metrics), SPD matrices (with affine-invariant, Log-Cholesky, and Log-Euclidean metrics), and others. For optimization, the library provides Riemannian adaptations of popular algorithms: stochastic gradient descent~\citep{bonnabel2013stochastic}, RMSprop~\citep{roy2018geometry}, and Adam~\citep{becigneul2018riemannian}. All optimizers support serialization for compatibility with TensorFlow's computational graph and model serving infrastructure.

Finally, the module \texttt{layers} exemplifies building blocks of neural networks, which can be used alongside with the native Keras~\citep{chollet2015keras} layers.

\section{Comparison to related software}

Several Python libraries provide Riemannian optimization capabilities for machine learning. We compare \texttt{tensorflow-riemopt} with existing tools, highlighting differences in framework integration and production readiness.

\texttt{Geoopt}~\citep{kochurov2020geoopt} is a research-oriented toolbox built on PyTorch~\citep{paszke2019pytorch}. It supports Riemannian SGD, adaptive optimization algorithms, and Markov chain Monte Carlo sampling methods. However, \texttt{Geoopt} does not provide integration with model serving and deployment infrastructure comparable to TensorFlow Extended.

\texttt{McTorch}~\citep{meghwanshi2018mctorch} is a manifold optimization library for deep learning that also leverages PyTorch capabilities. \texttt{McTorch} supports Riemannian variants of stochastic optimization algorithms, and also implements a collection of neural network layers with manifold-constrained parameters. \texttt{McTorch} shares the same limitations as \texttt{Geoopt} in its applicability for deployment in industrial settings.

\texttt{Pymanopt}~\citep{townsend2016pymanopt} is a Python package for Riemannian optimization that provides solvers including steepest descent, conjugate gradients, and trust region methods. It leverages JAX~\citep{bradbury2018jax}, PyTorch and TensorFlow backends for automatic differentiation. While \texttt{Pymanopt} is well-suited for general optimization problems, it lacks stochastic gradient descent algorithms necessary for large-scale neural network training and is not designed for production deployment.

\texttt{Geomstats}~\citep{miolane2020geomstats} package that focuses on research in differential geometry and education use cases, by providing low-level code that follows the \texttt{scikit-learn}~\citep{pedregosa2011scikit} semantics. \texttt{Geomstats} supports a broad list of manifolds and also provides a modular library of advanced differential geometry concepts, such as Levi-Civita connections and Christoffel symbols. \texttt{Geomstats} examples also contain a modified version of the Keras framework with support of Riemannian SGD algorithm; however, this fork lacks active maintenance and engineering support.

Finally, a recent benchmark by~\citet{axen2023manifolds} compared 
\texttt{tensorflow-riemopt} with nine other Riemannian optimization 
libraries across various manifolds and dimensions. The study found 
\texttt{tensorflow-riemopt} particularly effective for high-dimensional  manifolds, where it outperformed alternatives due to the optimized computational graph transformations. Further runtime benchmarks across related packages are provided in Appendix~\ref{appx:benchmarks}.

\section{Conclusion}
We present \texttt{tensorflow-riemopt}, a library for optimization on Riemannian manifolds in TensorFlow. The library provides infrastructure for deploying geometric machine learning methods such as hyperbolic neural networks, SPD matrix learning, and Grassmannian deep networks in production environments. By integrating manifold-constrained optimization with TensorFlow's ecosystem, \texttt{tensorflow-riemopt} enables practitioners to apply geometric methods without specialized expertise in differential geometry, while supporting researchers in developing novel algorithms through its low-level API. The library facilitates the transition from research prototypes to production deployment through compatibility with TensorFlow Extended and standard machine learning pipelines.

\bibliography{main}

\appendix
\section{Benchmarks}\label{appx:benchmarks}

Five open-source Riemannian optimization toolkits are benchmarked: \texttt{Geomstats}, \texttt{Pymanopt}, \texttt{Geoopt}, \texttt{tensorflow-riemopt}, and \texttt{McTorch}. Performance is evaluated on five manifolds commonly used in machine learning: Euclidean space $\mathbb{R}^n$, the Poincar\'{e} ball $\mathbb{H}^n$, the hypersphere $\mathbb{S}^{n-1}$, the special orthogonal group $\mathrm{SO}(n)$, and symmetric positive definite matrices $\mathcal{S}_{++}^n$ with affine-invariant metric. For each manifold, three fundamental operations are benchmarked: geodesic distance $d(x, y)$, the exponential map $\mathrm{Exp}_x(v)$, and the logarithmic map $\mathrm{Log}_x(y)$. Coverage varies across toolkits: \texttt{Geoopt} lacks $\mathrm{SO}(n)$ and \texttt{McTorch} supports only limited hyperbolic operations.

\begin{figure}[htp]
    \centering
    \begin{minipage}{0.3\textwidth}
        \centering
        \includegraphics[width=\linewidth]{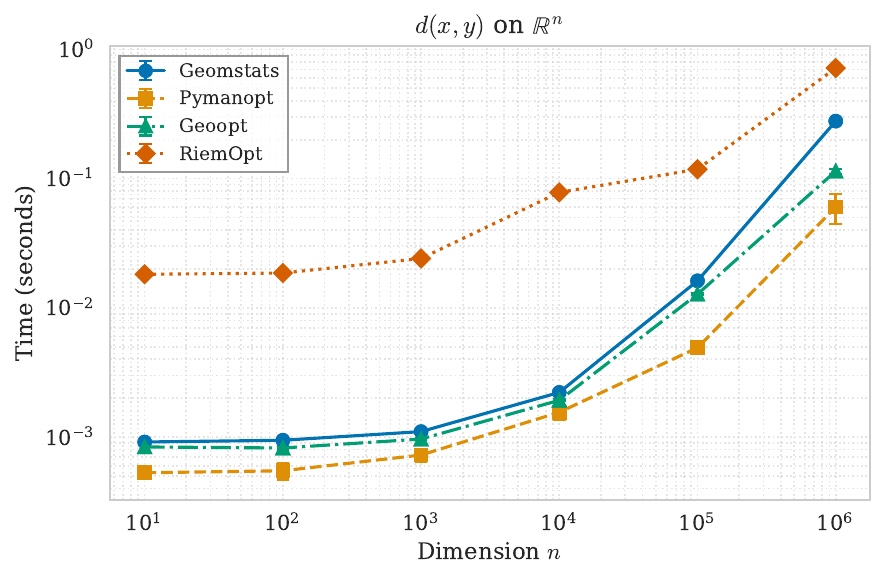}
    \end{minipage}%
    \hfill
    \begin{minipage}{0.3\textwidth}
        \centering
        \includegraphics[width=\linewidth]{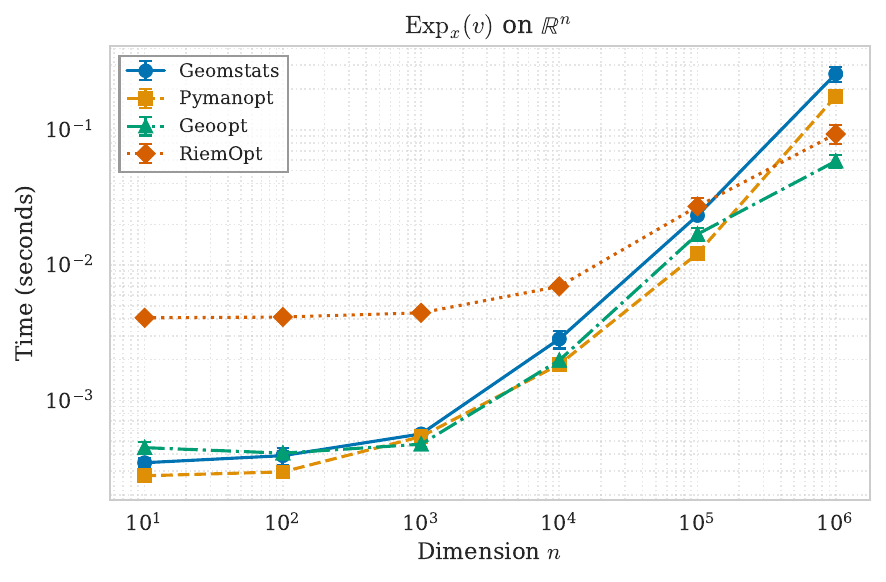}
    \end{minipage}%
    \hfill
    \begin{minipage}{0.3\textwidth}
        \centering
        \includegraphics[width=\linewidth]{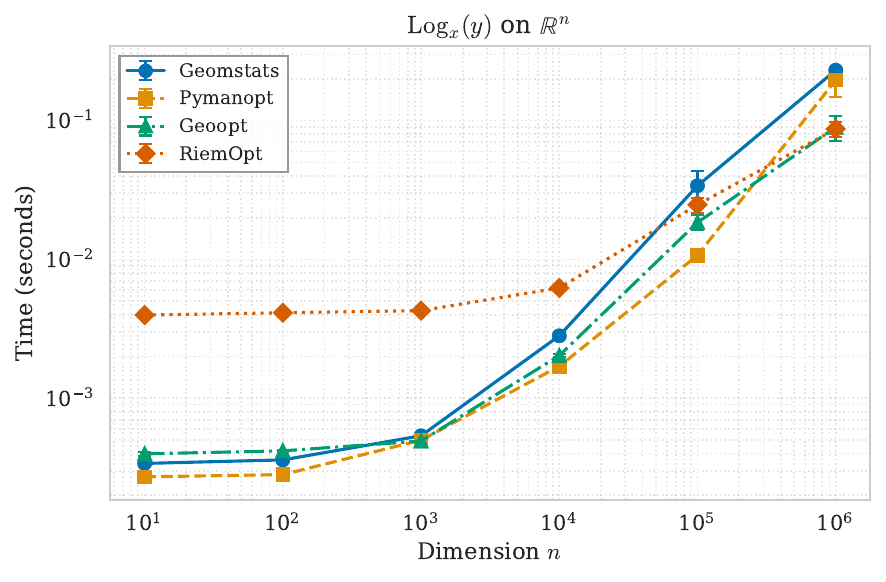}
    \end{minipage}
    \vspace{0.2cm}
    \begin{minipage}{0.3\textwidth}
        \centering
        \includegraphics[width=\linewidth]{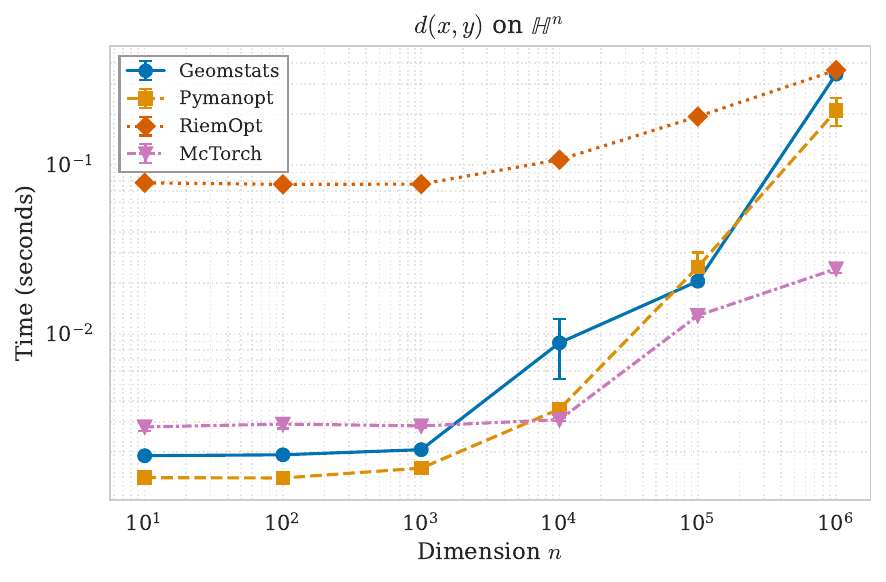}
    \end{minipage}%
    \hfill
    \begin{minipage}{0.3\textwidth}
        \centering
        \includegraphics[width=\linewidth]{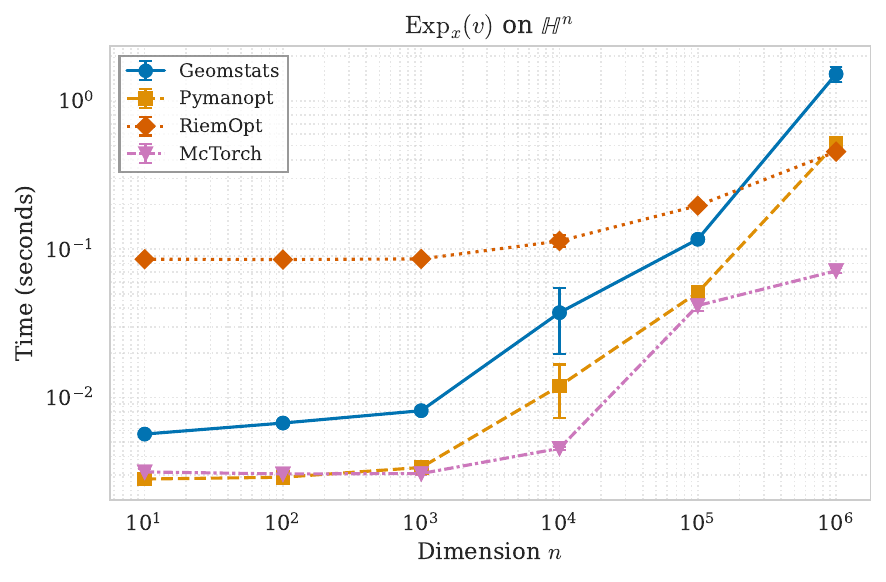}
    \end{minipage}%
    \hfill
    \begin{minipage}{0.3\textwidth}
        \centering
        \includegraphics[width=\linewidth]{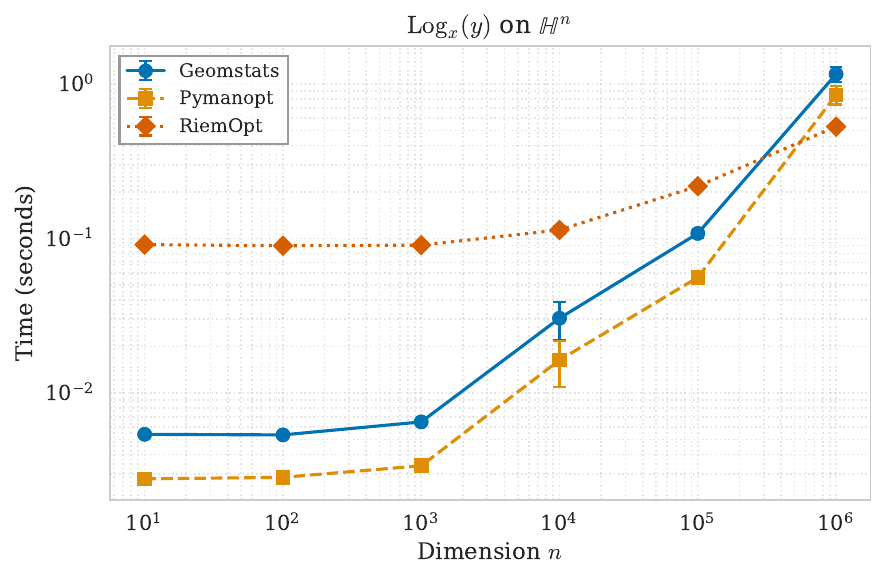}
    \end{minipage}
    \vspace{0.2cm}
    \begin{minipage}{0.3\textwidth}
        \centering
        \includegraphics[width=\linewidth]{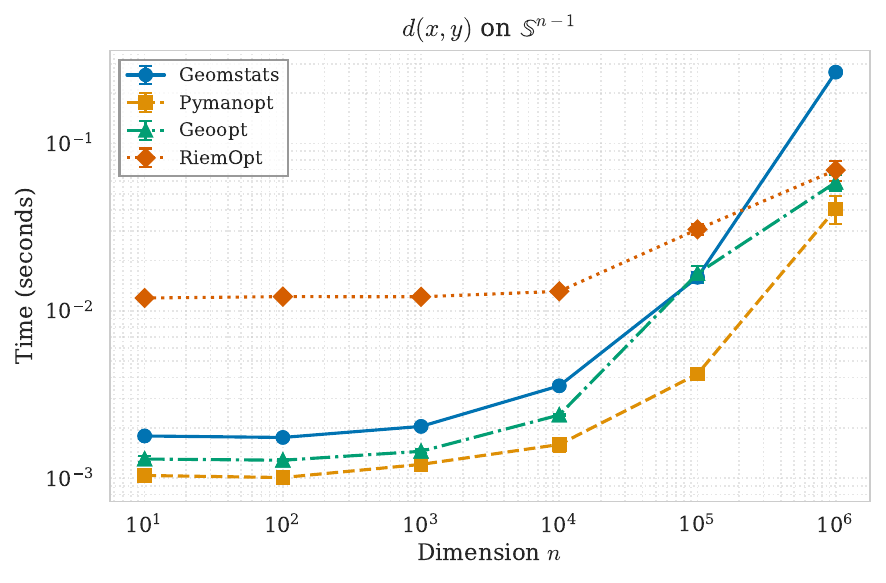}
    \end{minipage}%
    \hfill
    \begin{minipage}{0.3\textwidth}
        \centering
        \includegraphics[width=\linewidth]{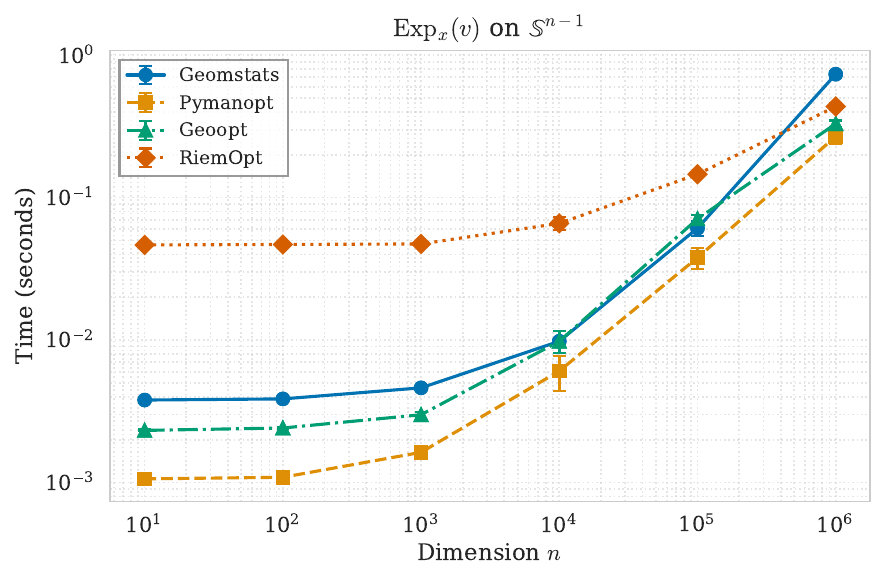}
    \end{minipage}%
    \hfill
    \begin{minipage}{0.3\textwidth}
        \centering
        \includegraphics[width=\linewidth]{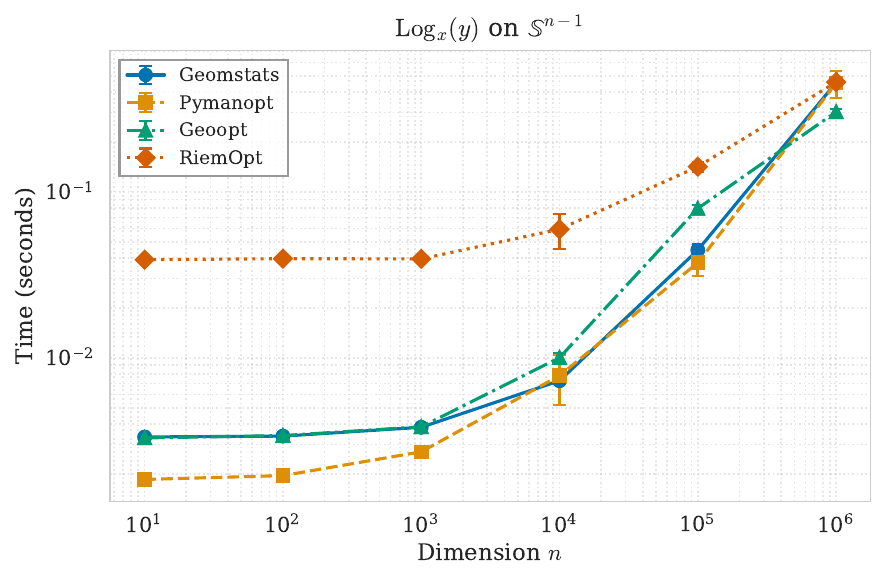}
    \end{minipage}
    \vspace{0.2cm}
    \begin{minipage}{0.3\textwidth}
        \centering
        \includegraphics[width=\linewidth]{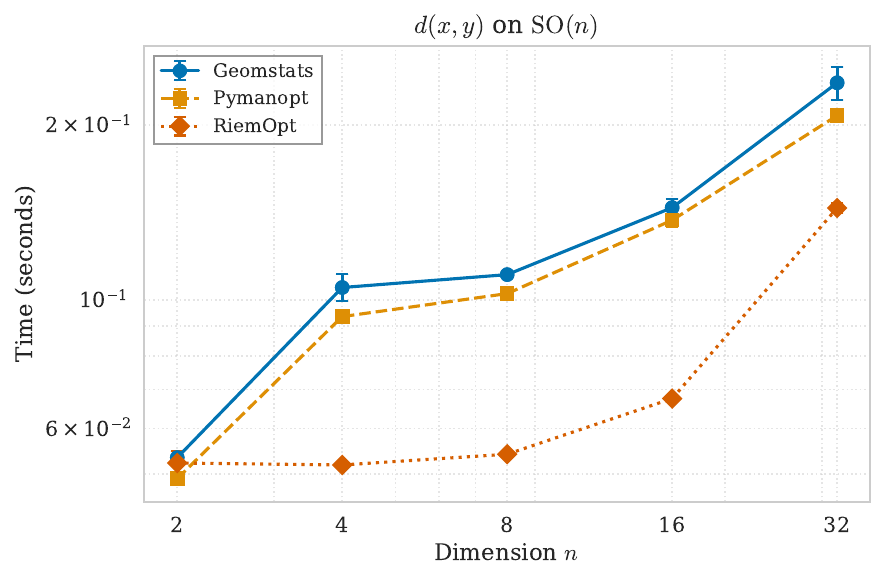}
    \end{minipage}%
    \hfill
    \begin{minipage}{0.3\textwidth}
        \centering
        \includegraphics[width=\linewidth]{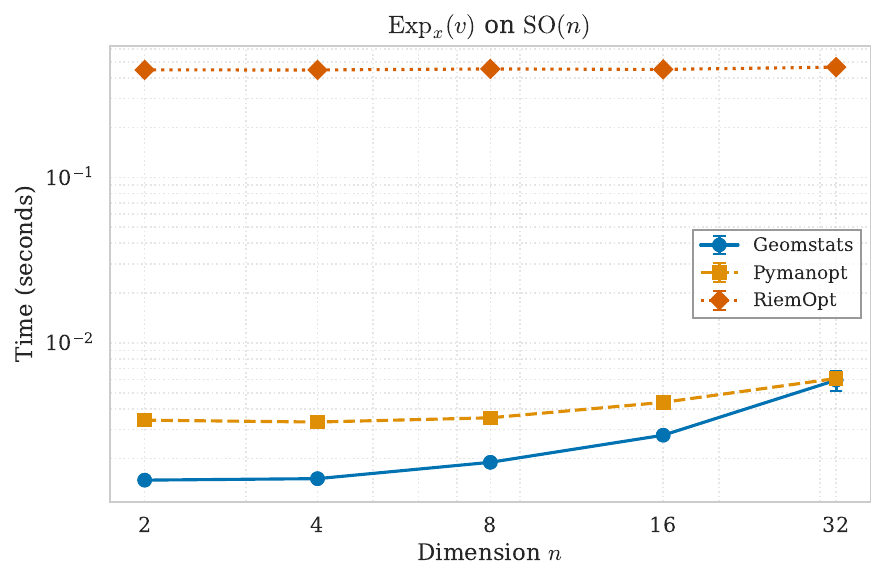}
    \end{minipage}%
    \hfill
    \begin{minipage}{0.3\textwidth}
        \centering
        \includegraphics[width=\linewidth]{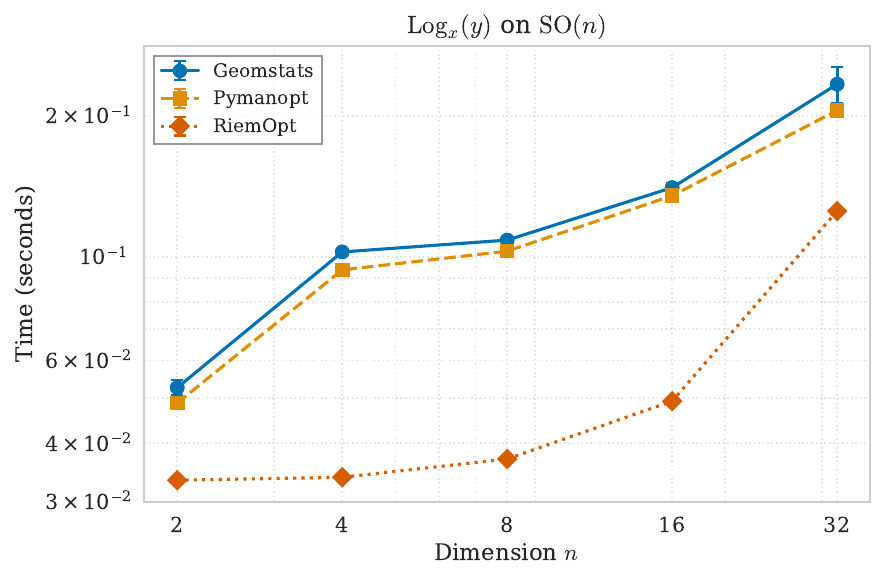}
    \end{minipage}
    \vspace{0.2cm}
    \begin{minipage}{0.3\textwidth}
        \centering
        \includegraphics[width=\linewidth]{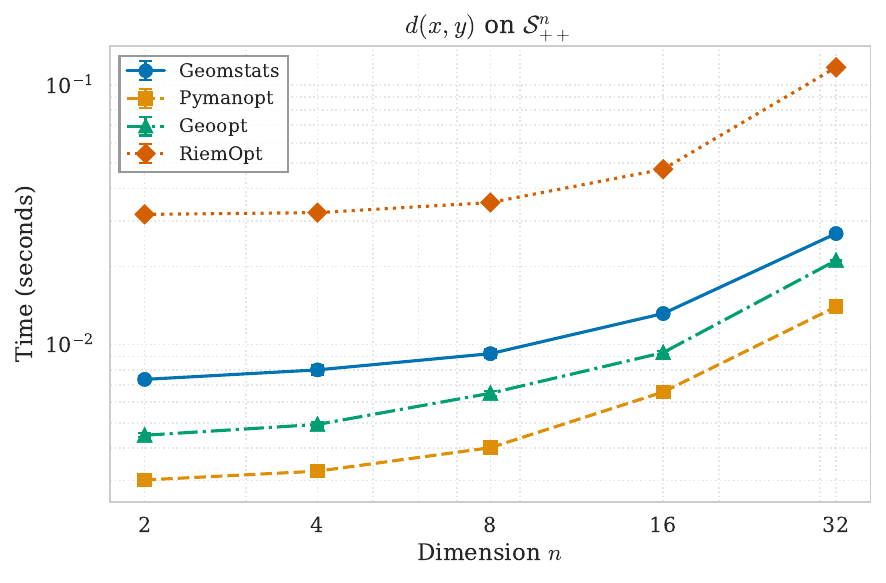}
    \end{minipage}%
    \hfill
    \begin{minipage}{0.3\textwidth}
        \centering
        \includegraphics[width=\linewidth]{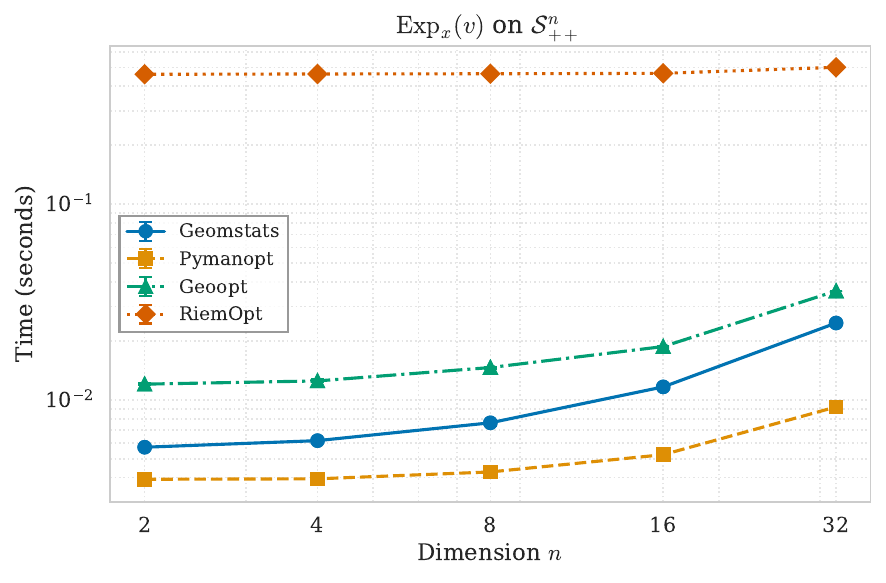}
    \end{minipage}%
    \hfill
    \begin{minipage}{0.3\textwidth}
        \centering
        \includegraphics[width=\linewidth]{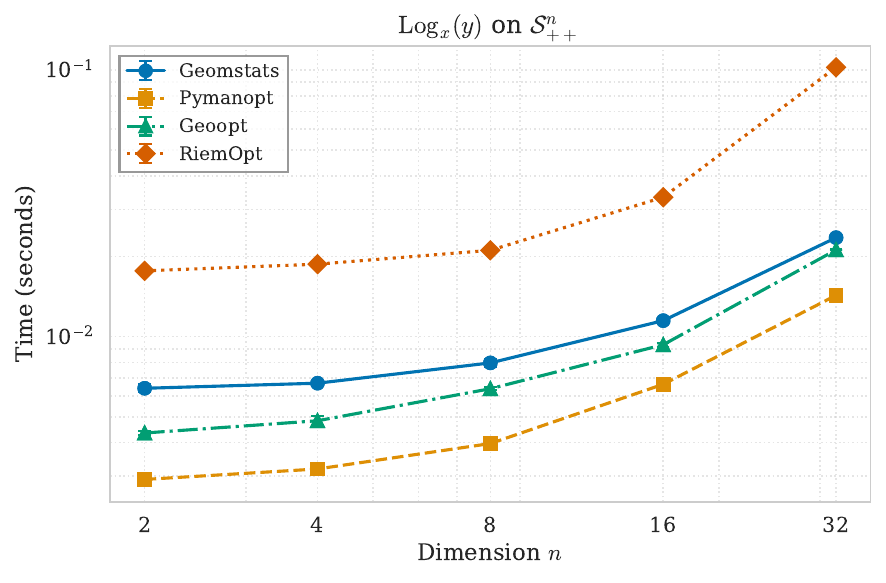}
    \end{minipage}
    \caption{Benchmarks for geodesic distance $d(x,y)$, exponential map $\mathrm{Exp}_x(v)$, and logarithmic map $\mathrm{Log}_x(y)$ across five manifolds: Euclidean space ($\mathbb{R}^n$), Poincar\'{e} ball ($\mathbb{H}^n$), hypersphere ($\mathbb{S}^{n-1}$), special orthogonal group ($\mathrm{SO}(n)$), and symmetric positive definite matrices ($\mathcal{S}^n_{++}$). In high-dimensional settings, \texttt{tensorflow-riemopt} performs comparably to or better than existing alternatives. The observed differences stem from implementation choices: exact geometric operations versus approximations such as retractions.}
    \label{fig:bench}
\end{figure}

For vector space manifolds, dimensions $n \in \{10, 10^2, 10^3, 10^4, 10^5, 10^6\}$ are tested, while for matrix manifolds $n \in \{2, 2^2, 2^3, 2^4, 2^5\}$ is used. Each benchmark generates 100 random point or point-tangent pairs, executes the operation with 5 warmup iterations followed by 10 timed iterations, and reports mean wall-clock time and standard deviation. The libraries support various backends: \texttt{Geomstats} (NumPy, PyTorch, Autograd), \texttt{Pymanopt} (NumPy, JAX, PyTorch, TensorFlow, Autograd), \texttt{Geoopt} and \texttt{McTorch} (PyTorch), and \texttt{tensorflow-riemopt} (TensorFlow). For fair comparison, all benchmarks run on an 8-core M2 CPU with a single backend per library. Complete results are depicted in Figure~\ref{fig:bench}.

\end{document}